\documentclass[10pt,journal,final]{IEEEtran}
\IEEEoverridecommandlockouts

\usepackage{amsfonts,amssymb}
\usepackage{ifpdf}
\usepackage{cite}
\usepackage{stfloats}
\usepackage{bm}
\usepackage{color,xcolor}
\usepackage{subfigure}
\ifCLASSINFOpdf
   \usepackage[pdftex]{graphicx}
   \graphicspath{{../pdf/}{../jpeg/}}
   \DeclareGraphicsExtensions{.pdf,.jpeg,.png}
\else
   \usepackage[dvips]{graphicx}

   \graphicspath{{../eps/}}

   \DeclareGraphicsExtensions{.eps}
\fi

\usepackage[cmex10]{amsmath}

\usepackage{algorithm}
\usepackage{algorithmic}
\ifCLASSOPTIONcompsoc
\else
  \usepackage[caption=false,font=footnotesize]{subfig}
\fi

\usepackage{makecell}
\begin{document}
\title{Weighted Sum-Rate Maximization for Multi-IRS Aided Cooperative Transmission}

\author{ Zhengfeng Li, Meng~Hua, Qingxia Wang, Qingheng Song

\thanks{Z. Li, Q. Wang, Q. Song are with the School of Electrical and Information Engineering, Huaihua University, Hunan, 418000, China (e-mail: \{Izf,  wqx\}@hhtc.edu.cn, qhsong@seu.edu.cn).}
\thanks{M. Hua is  with the School of Information Science and Engineering, Southeast University, Nanjing 210096, China (e-mail: mhua@seu.edu.cn).}
}
\maketitle
\begin{abstract}
This paper investigates multiple intelligent reflecting surfaces (IRSs) aided wireless network, where the IRSs are deployed to cooperatively assist communications between a multi-antenna base station (BS) and multiple single-antenna cell-edge users.  We aim at maximizing the weighted sum rate of all the cell-edge users by jointly optimizing the BS's transmit beamforming and IRS's phase shifts. Especially, the beamforming is  optimally  solved by the  Lagrangian method, and the phase shifts are obtained based on the Riemannian manifold  conjugate gradient (RMCG) method. Numerical results show that a significant throughput is improved with aid of multiple IRSs.
\end{abstract}
\section{Introduction}
An intelligent reflecting surface (IRS) has  emerged as a promising technique to increase the  throughput and spectral efficiency of wireless networks. Specifically, the IRS has a large number of reflective elements, each of which can independently control  the incident signal to change the signal propagation. Since each reflective element is a passive element consisted of  some low-cost printed dipoles, it  is a cost-effective and low-power consumption way to install it  on the room-ceilings, at buildings, even on lamp posts in the future \cite{wu2020towards}.

There have been many literatures paid attention on  integrating the IRS into the cellular network. Two main aspects are mostly be considered by the researchers, one is the channel estimation and the other is the  phase shift optimization. For the first aspect, different from the traditional channel estimation that the active device actively sends pilot signals  estimated by the terminal devices that can be  capable of  processing signal,  whereas the IRS is a passive device which cannot performing signal processing \cite{chen2019channel},\cite{you2019intelligent}. For the second  aspect, since the IRS reflects the  combined signal simultaneously, the phase shift matrix and BS transmit beamforming should be jointly optimized to increase the users' achievable rate \cite{liu2019joint,wu2019intelligent,pan2019intelligent}. Especially,  in \cite{wu2019intelligent}, an IRS-aided multiuser multiple input single-output  system was considered, and the phase shift matrix and BS transmit beamforming are jointly optimized by semidefinite relaxation and alternating optimization techniques. In \cite{pan2019intelligent}, the authors studied a simultaneous wireless information and power transfer system  aided by an IRS, and a dual decomposition and price-based method are used, which result in a low-complexity iterative algorithm.

However, the above works consider only one IRS, the multiple IRSs  has not been exploited. Especially in the cell edge region, the cell-edge users always suffer severe propagation, which leads to a poor communication service. Due to the limited IRS coverage, one IRS can not be satisfied with the users' high quality service requirements. To address this issue, we consider multiple IRSs deployed in a small cell, where  the IRSs and BS are managed by a central processing unit to coordinate transmission. Our goal is to maximize the weighted  sum rate (WSR) of all the cell-edge users by jointly optimizing the BS's transmit beamforming and each IRS's  phase shifts, subject to the BS transmit power limit. Since the resulting problem is a non-convex and unit-modulus constraint optimization problem, there is no standard convex technique to solve it. We equivalently transform the WSR problem into a weighted sum mean-square error (WMSE) problem, and a sub-optimal solution of the formulated problem is obtained based on the Lagrangian method and  Riemannian manifold  conjugate gradient (RMCG) method. Numerical results show that a significant throughput is improved with aid of the  IRSs and also show that the proposed iterative algorithm converges quite quickly.

\emph{Notations:~} Boldface lower-case and upper-case letter denote column vector and matrix, respectively. Transpose, conjugate, and transpose-conjugate operations are  denoted by ${\left(  \cdot  \right)^T}$, ${\left(  \cdot  \right)^*}$, and  ${\left(  \cdot  \right)^H}$, respectively. ${\left[ {\bf{Z}} \right]_{i,i}}$ represents the $i$th diagonal element of matrix $\bf Z$. ${\mathop{\rm Re}\nolimits} \left(  \cdot  \right)$ denotes the real part of a complex number. $ \odot $ is a Hadamard product operator. ${\mathbb E}\left(  \cdot  \right)$ is a expectation operator.

\begin{figure}[!t]
\centerline{\includegraphics[width=2.5in]{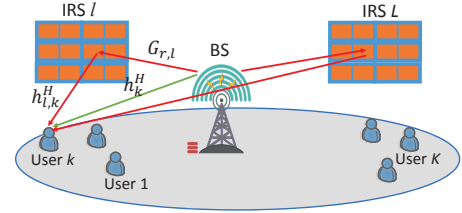}}
\caption{Multi-IRS aided Cooperative Transmission model} \label{fig1}
\end{figure}
\section{System Model And Problem Formulation}
Consider a multi-IRS aided  downlink network consisting of one base station (BS), $K$ single-antenna users, and $L$ intelligent reflecting surfaces (IRSs). We assume that the  BS is equipped with $N_t$ transmit antennas, and each IRS consists of $M$ phase  shifters. Let us denote the sets of users, phase shifters, and  IRSs as $\cal K$, $\cal M$ and $\cal L$, respectively. As shown in Fig.~\ref{fig1}, each user not only receives the signals  directly from the BS, but also receives the reflective signals from multiple IRSs. Note that the  signal reflected by multiple IRSs is ignored due to the severe propagation.

Mathematically, the transmitted signals by the  BS can be expressed as  ${\mathbf{x}} = \sum\limits_{k = 1}^K {{{\mathbf{w}}_k}{s_k}}$, where $s_k$ denotes  the desired  signal for user $k$ satisfying  ${\mathbb E}\left\{ {{s_k}s_k^H} \right\} = 1$ and ${\mathbb E}\left\{ {{s_i}s_j^H} \right\} = 0$ for $i \ne j$, and  ${{\bf{w}}_{k}} \in {{\mathbb C}^{{N_t} \times 1}}$  is BS transmit  beamforming  for the user $k$. Let ${\bf{h}}_{k}^H \in {{\mathbb  C}^{1 \times N_t}}$, ${\bf G}_{r,l} \in {{\mathbb C}^{M \times {N_t}}}$, and ${\bf h}_{l,k}^H \in {{\mathbb C}^{1 \times M}}$  respectively denote the complex equivalent baseband  channel vector between the $k$th user and the  BS, between the BS and the $l$-th IRS, and  between  the $l$-th IRS and the $k$-th user,  $\forall k \in {\cal K} $, $\forall l \in {\cal L}$. The received signal at the $k$th user is given by
\begin{align}
{{\bf{y}}_k} = {\bf{h}}_k^H{\bf{x}} + \sum\limits_{l = 1}^L {{\bf{h}}_{l,k}^H{{\bf{\Phi }}_l}{{\bf{G}}_{r,l}}{\bf{x}} + {n_k}}, \label{systemmodel1}
\end{align}
where ${{\bf{\Phi }}_l} = diag\left\{ {{e^{j\theta _l^1}}, \ldots ,{e^{j\theta _l^M}}} \right\}$ is a diagonal matrix that represents the adjustable phase shifts of the IRS $l$, wherein ${\theta _l^m}$ $(\forall m\in {\cal M}, \forall l\in {\cal L})$ is the $m$-th phase shifter at the $l$-th IRS, and $n_k$ is the received additive white Gaussian noise by the user $k$ with mean zero and variance $\sigma^2$. Note that here we assume that the amplitude of the reflection coefficient is maximized with 1. Substituting $\bf x$ into \eqref{systemmodel1}, we arrive at
\begin{align}
{{\bf{y}}_k} = {\bf{\bar h}}_k^H\sum\limits_{k = 1}^K {{{\bf{w}}_k}{s_k}}  + {n_k},\label{systemmodel2}
\end{align}
where ${\bf{\bar h}}_k^H = {\bf{h}}_k^H + \sum\limits_{l = 1}^L {{\bf{h}}_{l,k}^H{{\bf{\Phi }}_l}{{\bf{G}}_{r,l}}} $.
Accordingly,  the achievable data rate(nat/s/Hz) of the user $k$ is given by
\begin{align}
{R_k} = \log \left( {1 + {{{{\left| {{\bf{\bar h}}_k^H{{\bf{w}}_k}} \right|}^2}} \over {\sum\limits_{i \ne k}^K {{{\left| {{\bf{\bar h}}_k^H{{\bf{w}}_i}} \right|}^2} + {\sigma ^2}} }}} \right).\label{systemmodel3}
\end{align}
In this paper, we aim at maximizing the WSR of all users by jointly optimizing the BS transmit  beamforming $\{{\bf w}_k, \forall k\}$ and phase shift matrix $\{{{{\bf{\Phi }}_l}}, \forall l\}$, subject to the BS transmit power constraint. Define $\phi _l^m = {e^{j\theta _l^m}},\forall l,m$, we have ${{\bf{\Phi }}_l} = diag\left\{ {\phi _l^m, \ldots ,\phi _l^M} \right\}$. Then, the problem can be expressed as follow
\begin{align}
\left( {\rm{P}} \right)&\mathop {\max }\limits_{{{\bf{w}}_k},\phi _l^m} \sum\limits_{k = 1}^K {{\alpha _k}{R_k}} \notag\\
{\rm{s}}{\rm{.t}}{\rm{.}}~&\sum\limits_{k = 1}^K {\left\| {{{\bf{w}}_k}} \right\|_2^2}  \le {P_{\max }}, \label{const1}\\
&\left| {\phi _l^m} \right| = 1,\forall l,m, \label{const2}
\end{align}
where $\alpha _k\ge 0$ is a  weighting factor for the user $k$ with   a higher  value $\alpha _k$ representing  the higher priority for user $k$,  and  $P_{max}$ is the BS power limit.
\section{Proposed Low-Complexity Algorithm}
Problem $(\rm P)$ is challenging to solve since the non-convex rate expression \eqref{systemmodel3} in the  objective function and  unit-modulus constraint in \eqref{const2}. In the following, we first transform problem $(\rm P)$ into an equivalent weighted sum mean-square error (WMSE) problem, and then we  decouple the WMSE problem into several  sub-problems and alternately optimize the beamforming and phase shift matrix.


Specifically,  a decoder $u_k$ is applied at user $k$ to decode the desired signal $s_k$,  the estimated signal of the user $k$ is given by
\begin{align}
{\hat s_k} = u_k^H{y_k}. \label{P_exp1}
\end{align}
Under the independence assumption of signal $s_k$ and noise  $n_k$, the minimum MSE at user $k$ is given by
\begin{align}
{E_k} = & {{\mathbb E}_{s,n}}\left\{ {\left( {{{\hat s}_k} - {s_k}} \right){{\left( {{{\hat s}_k} - {s_k}} \right)}^H}} \right\}\notag\\
 = &u_k^H\left( {{\bf{\bar h}}_k^H\left( {\sum\limits_{k = 1}^K {{{\bf{w}}_k}{\bf{w}}_k^H} } \right){{{\bf{\bar h}}}_k} + {\sigma ^2}} \right){u_k} - \notag \\
 &u_k^H{\bf{\bar h}}_k^H{{\bf{w}}_k} - {\bf{w}}_k^H{{{\bf{\bar h}}}_k}{u_k} + 1. \label{P_exp2}
\end{align}
\textbf{\emph{Lemma 1:} } The weighted sum-rate maximization problem is equivalent to the WMSE problem $(\rm P1)$, which is given by
\begin{align}
\left( {{\rm{P}}1} \right)&\mathop {\max }\limits_{{u_k},{q_k>0},{{\bf{w}}_k},\phi _l^m} \sum\limits_{k = 1}^K {{\alpha _k}\left( {\log \left( {{q_k}} \right) - {q_k}{E_k} + 1} \right)} \notag \\
&{\rm s.t.}~ \eqref{const1},~\eqref{const2}.\notag
\end{align}
\hspace*{\parindent}\textit{Proof}: A brief  proof is given  in \textbf{Remark 1} in the later Subsection, and the detailed proof can be referred to  Theorem 1 in \cite{shi2011iteratively}.

Despite $(\rm P1)$ introduces two additional variables $\{u_k, \forall k\}$ and  $\{q_k, \forall k\}$, $(\rm P1)$ is much easier  to solve by using  the alternating optimization method as follows.
\subsection{Decoder $u_k$ optimization} \label{subsection1}
In this subsection, we  optimize $\{u_k\}$ while fixing $\{q_k\}$, phase shift $\{\phi _l^m\}$, and beamforming vector  $\{{{{\bf{w}}_k}}\}$. The simplified problem is given by
\begin{align}
\left( {{\rm{P}}1.1} \right)\mathop {\max }\limits_{{u_k}} \sum\limits_{k = 1}^K {{\alpha _k}\left( {\log \left( {{q_k}} \right) - {q_k}{E_k} + 1} \right)}.  \notag
\end{align}
Define ${f_k} = \log \left( {{q_k}} \right) - {q_k}{E_k} + 1$, and substitute \eqref{P_exp2} into $f_k$, we arrive at \eqref{P1-1_exp1}.
\newcounter{mytempeqncnt0}
\begin{figure*}
\normalsize
\setcounter{mytempeqncnt0}{\value{equation}}
\begin{align}
{f_k} = \log \left( {{q_k}} \right) - {q_k}\left( {u_k^H\left( {{\bf{\bar h}}_k^H\left( {\sum\limits_{k = 1}^K {{{\bf{w}}_k}{\bf{w}}_k^H} } \right){{{\bf{\bar h}}}_k} + {\sigma ^2}} \right){u_k} - u_k^H{\bf{\bar h}}_k^H{{\bf{w}}_k} - {\bf{w}}_k^H{{{\bf{\bar h}}}_k}{u_k} + 1} \right)+1. \label{P1-1_exp1}
\end{align}
\hrulefill 
\vspace*{4pt} 
\end{figure*}
It can be easily seen that \eqref{P1-1_exp1} is concave with respective to (w.r.t.) $u_k$, which  thus can be optimally solved by setting the  first-order derivative of $f_k$ w.r.t. $u_k$ to zero. We thus have
\begin{align}
u_k^{opt} = {{{\bf{\bar h}}_k^H{{\bf{w}}_k}} \over {\sum\limits_{j = 1}^K {{{\left| {{\bf{\bar h}}_k^H{{\bf{w}}_j}} \right|}^2} + {\sigma ^2}} }}.
\end{align}
\subsection{Optimal solution  $q_k$} \label{subsection2}
For any given $\{u_k\}$, phase shift $\{\phi _l^m\}$, and beamforming vector  $\{{{{\bf{w}}_k}}\}$, the optimal solution $q_k$ can be obtained by solving the following problem
\begin{align}
\left( {{\rm{P}}1.2} \right)\mathop {\max }\limits_{{q_k>0}} \sum\limits_{k = 1}^K {{\alpha _k}\left( {\log \left( {{q_k}} \right) - {q_k}{E_k} + 1} \right)}.  \notag
\end{align}
We can also see that $q_k$ is concave w.r.t. $f_k$ in $(\rm P1.2)$, the optimal solution $q_k$ can be easily solved by taking the first-order derivative of \eqref{P1-1_exp1}  w.r.t. $q_k$, we then have
\begin{align}
q_k^{opt} = E_k^{ - 1}.
\end{align}

\textbf{\emph{Remark 1:}} Based on the optimal solutions $u_k^{opt}$ and $q_k^{opt}$  obtained from \eqref{subsection1} and \eqref{subsection2}, substitute $u_k^{opt}$ and $q_k^{opt}$ into \eqref{P1-1_exp1}, we arrive at
\begin{align}
{f_k} &= \log \left( {E_k^{ - 1}} \right)\notag\\
&= \log \left( {1 + {{{{\left| {{\bf{\bar h}}_k^H{{\bf{w}}_k}} \right|}^2}} \over {\sum\limits_{i \ne k}^K {{{\left| {{\bf{\bar h}}_k^H{{\bf{w}}_i}} \right|}^2} + {\sigma ^2}} }}} \right) \overset{\triangle }{=} {R_k}.
\end{align}
This result shows the equivalence between  problem $(\rm P)$ and $(\rm P1)$.
\subsection{Lagrangian method for beamforming  optimization} \label{subsection3}
In this subsection, the optimal beamforming vector $\{{\bf w}_k\}$  is obtained by applying the Lagrangian method  \cite{boyd2004convex}. With the fixed variables  $\{q_k\}$, $\{u_k\}$ and  phase shift $\{\phi _l^m\}$, and drop the irrelevant terms with ${\bf w}_k$, the beamforming optimization problem can be simplified as
\begin{align}
\left( {{\rm{P}}1.3} \right)&\mathop {\min }\limits_{{{\bf{w}}_k}} \sum\limits_{k = 1}^K {{\alpha _k}{q_k}{E_k}} \notag\\
{\rm s.t.}&~\eqref{const1}. \notag
\end{align}
It can be easily checked that the objective function and constraint in $(\rm P1.3)$ are all convex, which can be efficiently solved by the convex tools such CVX \cite{cvx}. To reduce the computational complexity generally solved  by CVX,  we obtain a globally optimal solution to $(\rm P1.3)$ with a much lower  complexity based on the Lagrangian method. To this end, we first introduce a non-negative slack variable $\lambda$ associated with constraint \eqref{const1}, the Lagrangian function of problem $(\rm P1.3)$ is thus given by
\begin{align}
 \hat {\cal L}\left( {{{\bf{w}}_k},\lambda } \right) = \sum\limits_{k = 1}^K {{\alpha _k}{q_k}{E_k}}  + \lambda \left( {\sum\limits_{k = 1}^K {\left\| {{{\bf{w}}_k}} \right\|_2^2}  - {P_{\max }}} \right). \label{P1-3_exp1}
\end{align}
With any given $\lambda$, the optimal solution ${\bf w}_k$ to minimize \eqref{P1-3_exp1} can be obtained by directly setting  its first-order derivative of  $\hat {\cal L}\left( {{{\bf{w}}_k} } \right)$ w.r.t. ${\bf w}_k$ to zero, we have
\begin{align}
{\bf{w}}_k^{opt}(\lambda) = {\left( {\sum\limits_{j = 1}^K {{\alpha _j}{q_j}{{{\bf{\bar h}}}_j}{u_j}u_j^H{\bf{\bar h}}_j^H + \lambda {\bf I}} } \right)^{ - 1}}{\alpha _k}{q_k}{{{\bf{\bar h}}}_k}{u_k}, \label{P1-3_exp2}
\end{align}
where ${\bf{I}} \in {{\mathbb C}^{{N_t} \times {N_t}}}$ is an identity matrix. Define ${\bf{H}} = \sum\limits_{j = 1}^K {{\alpha _j}{q_j}{{{\bf{\bar h}}}_j}{u_j}u_j^H{\bf{\bar h}}_j^H}$. Since ${\bf{H}}$ is a positive semi-definite matrix, we assume that the rank of ${\bf{H}}$ as $N$ $(N<=N_t)$, it thus  can be decomposed as
\begin{align}
{\bf{H}} = \left[ {{{\bf{F}}_1}{\kern 1pt} {\kern 1pt} {\kern 1pt} {\kern 1pt} {\kern 1pt} {\kern 1pt} {{\bf{F}}_2}} \right]diag\left( {{{\bf{\Sigma }}_1},{{\bf{\Sigma }}_2}} \right){\left[ {{{\bf{F}}_1}{\kern 1pt} {\kern 1pt} {\kern 1pt} {\kern 1pt} {\kern 1pt} {\kern 1pt} {{\bf{F}}_2}} \right]^H}, \label{P1-3_exp3}
\end{align}
where ${\bf F}_1$ is the first $N$ singular vectors corresponding to the $N$ positive  eigenvalues in diagonal matrix ${{{\bf{\Sigma }}_1}}$, and ${\bf F}_2$  is the remaining $N_t-N$ singular vectors corresponding to the $N_t-N$  zero  eigenvalues in ${{{\bf{\Sigma }}_2}}$. We thus can simplify \eqref{P1-3_exp3} as
\begin{align}
{\bf{H}} = {{\bf{F}}_1}{{\bf{\Sigma }}_1}{\kern 1pt} {\bf{F}}_1^H.\label{P1-3_exp4}
\end{align}

With  \eqref{P1-3_exp2} and \eqref{P1-3_exp4}, we have $g\left( \lambda  \right)$, which is expressed in  \eqref{P1-3_exp5}, where ${{\varepsilon _i}}$ is the $i$th diagonal element in ${{{\bf{\Sigma }}_1}}$, and ${{\bf{Z}}_k} = {\bf{F}}_1^H{{{\bf{\bar h}}}_k}{u_k}u_k^H{\bf{\bar h}}_k^H{{\bf{F}}_1}$.
\newcounter{mytempeqncnt1}
\begin{figure*}
\normalsize
\setcounter{mytempeqncnt1}{\value{equation}}
\begin{small}
\begin{align}
g\left( \lambda  \right) &= \sum\limits_{k = 1}^K {\left\| {{{\bf{w}}_k}} \right\|_2^2}  = \sum\limits_{k = 1}^K {{\rm{Tr}}\left( {{{\bf{F}}_1}{{\left( {{{\bf{\Sigma }}_1}{\kern 1pt}  + \lambda {\bf{I}}} \right)}^{ - 1}}{\bf{F}}_1^H{\alpha _k}{q_k}{{{\bf{\bar h}}}_k}{u_k}u_k^H{\bf{\bar h}}_k^H{q_k}{\alpha _k}{{\bf{F}}_1}{{\left( {{{\bf{\Sigma }}_1}{\kern 1pt}  + \lambda {\bf{I}}} \right)}^{ - 1}}{\bf{F}}_1^H} \right)}\notag\\
&= \sum\limits_{k = 1}^K {{{\left| {{\alpha _k}} \right|}^2}{{\left| {{q_k}} \right|}^2}} {\rm{Tr}}\left( {{{\left( {{{\bf{\Sigma }}_1}{\kern 1pt}  + \lambda {\bf{I}}} \right)}^{ - 2}}{\bf{F}}_1^H{{{\bf{\bar h}}}_k}{u_k}u_k^H{\bf{\bar h}}_k^H{{\bf{F}}_1}} \right)\notag\\
& = \sum\limits_{k = 1}^K {{{\left| {{\alpha _k}} \right|}^2}{{\left| {{q_k}} \right|}^2}} \sum\limits_{i = 1}^N {{{{{\left[ {{{\bf{Z}}_k}} \right]}_{i,i}}} \over {{{\left( {{\varepsilon _i} + \lambda } \right)}^2}}}},  \label{P1-3_exp5}
\end{align}
\end{small}
\hrulefill 
\vspace*{4pt} 
\end{figure*}
The optimal $\lambda$ must be chosen for satisfying the complementary slackness condition for BS power constraint as follow
\begin{align}
\lambda \left( {g\left( \lambda  \right) - {P_{\max }}} \right) = 0.
\end{align}
As can be seen in \eqref{P1-3_exp5}, ${g\left( \lambda  \right)}$ is a decreasing function of $\lambda$. As a consequence, if $g(0)\le P_{\rm max}$, the optimal  beamforming vector is ${\bf{w}}_k^{opt}(0) = {\left( {\sum\limits_{j = 1}^K {{\alpha _j}{q_j}{{{\bf{\bar h}}}_j}{u_j}u_j^H{\bf{\bar h}}_j^H} } \right)^{ - 1}}{\alpha _k}{q_k}{{{\bf{\bar h}}}_k}{u_k}$. Otherwise, if $g(0)>P_{\rm max}$, the optimal $\lambda^{opt}$ can be found via bisection based search  method to ensure $g\left( {{\lambda ^{opt}}} \right) - {P_{\max }} = 0$. Then, the optimal beamforming vector  can be obtained as  ${\bf{w}}_k^{opt}\left( {{\lambda ^{opt}}} \right) = {\left( {\sum\limits_{j = 1}^K {{\alpha _j}{q_j}{{{\bf{\bar h}}}_j}{u_j}u_j^H{\bf{\bar h}}_j^H + {\lambda ^{opt}}{\bf{I}}} } \right)^{ - 1}}{\alpha _k}{q_k}{{{\bf{\bar h}}}_k}{u_k}$. To reduce the search range  $\left[ {{\lambda _{\min }}{\kern 1pt} {\kern 1pt} {\kern 1pt} {\kern 1pt} {\lambda _{\max }}} \right]$, the initial lower bound of $\lambda$ is set as $\lambda_{\rm min}=0$, and the initial upper  bound of $\lambda$ is calculated as follow
\begin{align}
g\left( \lambda  \right) \le \sum\limits_{k = 1}^K {{{\left| {{\alpha _k}} \right|}^2}{{\left| {{q_k}} \right|}^2}} \sum\limits_{i = 1}^N {{{{{\left[ {{{\bf{Z}}_k}} \right]}_{i,i}}} \over {\lambda _{\max }^2}} \overset{\triangle} {=} } {P_{\max }},\notag\\
\Rightarrow {\lambda _{\max }} = \sqrt {{{\sum\limits_{k = 1}^K {{{\left| {{\alpha _k}} \right|}^2}{{\left| {{q_k}} \right|}^2}\sum\limits_{i = 1}^N {{{\left[ {{{\bf{Z}}_k}} \right]}_{i,i}}} } } \over {{P_{\max }}}}}.
\end{align}
\subsection{ RMCG method for phase shift optimization} \label{subsection4}
In this subsection, with fixed $\{q_k\}$, $\{u_k\}$, and beamforming vector $\{{\bf w}_k\}$, we consider the phase shift optimization problem, which is given by
\begin{align}
\left( {{\rm{P}}1.4} \right)&\mathop {\min }\limits_{\phi _l^m} \sum\limits_{k = 1}^K {{\alpha _k}{q_k}{E_k}}\notag\\
{\rm s.t.}&~\eqref{const2}. \notag
\end{align}
Problem $(\rm P1.4)$ is non-convex due to the  unit-modulus constraint in \eqref{const2}, the globally optimal solution is hard to achieve in general. In order to develop an efficient algorithm to solve $(\rm P1.4)$, we develop  a  Riemannian manifold  conjugate gradient (RMCG) method, which guarantees at least a locally optimal solution \cite{alhujaili2019transmit}. Substituting ${\bf{\bar h}}_k^H$ into $E_k$, we can rewrite   ${ E}_k$ as
\begin{small}
\begin{align}
&{{ E}_k} = \sum\limits_{i = 1}^L {\sum\limits_{j =1 }^L {{\rm{Tr}}\left( {{\bf{\Phi }}_i^H{{\bf{A}}_{i,j,k}}{{\bf{\Phi }}_j}{{{\bf{\bar E}}}_{i,j}}} \right)} }  + \sum\limits_{l = 1}^L {{\rm{Tr}}\left( {{\bf{\Phi }}_l^H\left( {{{\bf D}_{l,k}} - {{\bf B}_{l,k}}} \right)} \right)}\notag\\
&+\sum\limits_{l = 1}^L {{\rm{Tr}}\left( {{{\bf{\Phi }}_l}{{\left( {{{\bf D}_{l,k}} - {{\bf B}_{l,k}}} \right)}^H}} \right)}+{{c_k} - {e_k} - e_k^H + 1}, \label{P1-3_exp6}
\end{align}
\end{small}
where ${{\bf{A}}_{i,j,k}} = {{\bf{h}}_{i,k}}{u_k}u_k^H{\bf{h}}_{j,k}^H$, ${{\bf{B}}_{l,k}} = {{\bf{h}}_{l,k}}{u_k}{\bf{w}}_k^H{\bf{G}}_{r,l}^H$, ${{\bf{D}}_{l,k}} = {{\bf{h}}_{l,k}}{u_k}u_k^H{\bf{h}}_k^H\left( {\sum\limits_{k = 1}^K {{{\bf{w}}_k}{\bf{w}}_k^H} } \right){\bf{G}}_{r,l}^H$, ${{{\bf{\bar E}}}_{i,j}} = {{\bf{G}}_{r,j}}\left( {\sum\limits_{k = 1}^K {{{\bf{w}}_k}{\bf{w}}_k^H} } \right){\bf{G}}_{r,i}^H$, ${c_k} = u_k^H\left( {{\bf{h}}_k^H\left( {\sum\limits_{k = 1}^K {{{\bf{w}}_k}{\bf{w}}_k^H} } \right){{\bf{h}}_k} + {\sigma ^2}} \right){u_k}$, and ${e_k} = {\bf{w}}_k^H{{\bf{h}}_k}{u_k}$.
Define vector ${{\bf v}_l} = {\left[ {\phi _l^1, \ldots ,\phi _l^M} \right]^T}$ for $\forall l$, by dropping the constant terms $c_k$ and $e_k$ irrespective to variable ${\phi _l^m}$ in  $E_k$,   $(\rm P1.4)$ can be equivalently written as
\begin{align}
\left( {{\rm{\bar P}}1.4} \right)&\mathop {\min }\limits_{\phi _l^m} \sum\limits_{i = 1}^L {\sum\limits_{j = 1}^L {{\bf{v}}_i^H} } {{\bf{J}}_{i,j}}{{\bf{v}}_j} + \sum\limits_{l = 1}^L {{\bf{v}}_l^H{{\bf{z}}_l}}  + \sum\limits_{l = 1}^L {{\bf{z}}_l^H{{\bf{v}}_l}} \notag\\
{\rm s.t.}&~\eqref{const2}. \notag
\end{align}
where ${{\bf{J}}_{i,j}} = \left( {\sum\limits_{k = 1}^K {{a_k}{q_k}{{\bf{A}}_{i,j,k}}} } \right) \odot {\bf{\bar E}}_{i,j}^T$, ${{\bf{z}}_l} = {\left[ {\sum\limits_{k = 1}^K {{a_k}{q_k}{{\left[ {{{\bf{D}}_{l,k}} - {{\bf{B}}_{l,k}}} \right]}_{1,1}}, \ldots ,\sum\limits_{k = 1}^K {{a_k}{q_k}{{\left[ {{{\bf{D}}_{l,k}} - {{\bf{B}}_{l,k}}} \right]}_{M,M}}} } } \right]^T}$.
Define ${\bf{\hat v}} = \left[ {{\bf{v}}_1^T, \ldots ,{\bf{v}}_L^T} \right]^T$, ${\bf{z}} = {\left[ {{\bf{z}}_1^T, \ldots ,{\bf{z}}_L^T} \right]^T}$, and ${{\bf{\hat J}}}=\left(                 
  \begin{array}{ccc}   
    {\bf J}_{1,1} &  \cdots  & {\bf J}_{1,L} \\  
     \vdots  &  \cdots  &  \vdots \\
      {\bf J}_{L,1} &  \cdots  & {\bf J}_{L,L}\\
  \end{array}
\right)$. We can equivalently rewrite $\left( {{\rm{\bar P}}1.4} \right)$ in a more simplified form as follow
\begin{align}
\left( {{\rm{\hat P}}1.4} \right)&\mathop {\min }\limits_{{\bf{\hat v}}} \hat f\left( {{\bf{\hat v}}} \right) = {{{\bf{\hat v}}}^H}\left( {{\bf{\hat J}} + \omega {\bf{I}}} \right){\bf{\hat v}} + {{{\bf{\hat v}}}^H}{\bf{z}} + {{\bf{z}}^H}{\bf{\hat v}}\notag\\
{\rm s.t.}&~\eqref{const2}. \notag
\end{align}
where $\omega$ is a auxiliary constant which can be used to speed up the convergence of the proposed RMCG method \cite{alhujaili2019transmit}, \cite{pan2019multicell}. Note that ${{{\bf{\hat v}}}^H}\omega {\bf{I\hat v}} = \omega ML$, which indicates   it will not change the optimal
solution to $\left( {{\rm{\bar P}}1.4} \right)$.
We first define the manifold space for constraint \eqref{const2} in  $(\rm {\hat P}1.4)$ as
\begin{align}
{{\cal S}^{ML}} = \left\{ {{\bf{\hat v}} \in {{\mathbb C}^{ML \times 1}}:\left| {\phi _1^1} \right| =  \cdots  = \left| {\phi _1^M} \right| =  \cdots \left| {\phi _L^M} \right| = 1} \right\},
\end{align}
where ${\cal S} = \left\{ {{{{\bf{\hat v}}}_{lm}} \in {\mathbb C}:\left| {\phi _l^m} \right| = 1} \right\}$ is a complex circle, which can be regarded as a sub-manifold of ${\cal S}^{ML}$. Precisely, $(\rm {\hat P}1.4)$ can be solved iteratively by preforming the following steps at each iteration $r$ \cite{chen2017manifold}: 
1) We first find the gradient in Euclidean space ${{\bm{\eta }}^r} =  - {\nabla _{\phi _l^m}}\hat f\left( {{{{\bf{\hat v}}}^r}} \right) =  - 2\left( {{\bf{\hat J}} + \omega {\bf{I}}} \right){{{\bf{\hat v}}}^r} - 2{\bf{z}}$. 
2) We then compute the Riemannian gradient of $\hat f\left( {{{{\bf{\hat v}}}^r}} \right)$ at point ${{{{\bf{\hat v}}}^r}}$ via projecting onto the tangent space ${{\cal T}_{{{{\bf{\hat v}}}^r}}}{{\cal S}^{ML}}$, the Riemannian gradient is  then given by ${{\cal T}_{{{{\bf{\hat v}}}^r}}}{{\cal S}^{ML}} = {{\bm {\eta}} ^r} - {\mathop{\rm Re}\nolimits} \left\{ {{{\bm{\eta }}^{r * }} \odot {{{\bf{\hat v}}}^r}} \right\} \odot {{{\bf{\hat v}}}^r}$. 3) Then, update the current 
value of ${{{{\bf{\hat v}}}^r}}$ onto the ${{\cal T}_{{{{\bf{\hat v}}}^r}}}{{\cal S}^{ML}}$, the update is given by
${{{\bf{\hat v}}}^{r + 1}} = {{{\bf{\hat v}}}^r} + \zeta {T_{{{{\bf{\hat v}}}^r}}}{{\cal S}^{ML}}$, where $\zeta$ is a conjugate parameter. 4)  We then  map  ${{{\bf{\hat v}}}^{r + 1}}$ into the manifold space ${{\cal S}^{ML}}$ by performing retraction operator, which is given by ${{{\bf{\hat v}}}^{r + 1}} = {{{\bf{\hat v}}}^{r + 1}} \odot {1 \over {\left| {{{{\bf{\hat v}}}^{r + 1}}} \right|}}$.
\subsection{Overall algorithm and complexity analysis}
Based on the solutions  to  sub-problems, an iterative algorithm is performed to alternately optimize the four sub-problems until the fractional increase of the objective value  less than a predefined value. It should be pointed out  that the complexity of this iterative algorithm is very low. The main complexity of  proposed algorithm mainly lies in \eqref{subsection3} and \eqref{subsection4}. In \eqref{subsection3}, the main complexity includes  three parts. First, the complexity of decomposing $\bf H$ in \eqref{P1-3_exp3} is ${\cal O}\left( {{{\left( {{N_t}} \right)}^3}} \right)$. Second, the complexity of searching optimal  $\lambda$ is given by ${\cal O}\left( {{{\log }_2}\left( {{{{\lambda _{\max }} - {\lambda _{\min }}} \over \tau }} \right)} \right)$, where $\tau$ is a tolerant  value. Last, the complexity of calculating  optimal ${\bf w}_k$ involving inverse operator is ${\cal O}\left( {{{\left( {{N_t}} \right)}^3}} \right)$.
In \eqref{subsection4}, the main complexity lies in  calculating the Euclidean gradient ${{\bm{\eta }}^r}$ in step 1, which is given by  ${\cal O}\left( {{{\left( {ML} \right)}^2}} \right)$. Then, the total complexity of the proposed iterative algorithm is ${\cal O}\left( {{\kappa _1}\left( {{\kappa _2}{{\left( {ML} \right)}^2} + {{\left( {{N_t}} \right)}^3} + {{\log }_2}\left( {{{{\lambda _{\max }} - {\lambda _{\min }}} \over \tau }} \right)} \right)} \right)$, where $\kappa_1$ and $\kappa_2$ respectively represent the  number of iterations  required by the overall algorithm and  RMCG method.
\section{Numerical  results}
In this section, numerical simulations are provided to evaluate the performance of the joint optimization  of  BS transmit beamforming and IRSs' phase shifts. The BS is located at $(0,0)$ with radius $300\rm m$ and  height  $10\rm m$. We consider 4 IRSs, which are respectively located at $(300 \rm m, 0)$, $(0, 300 \rm m)$, $(-300 \rm m, 0)$, and $(0, -300 \rm m)$ with height $10\rm m$. There are 8 users, each two of them are uniformly and randomly placed in a circle centered at each   location of IRS with radius $30\rm m$. The large-scale path is denoted as  ${L_{loss}} = {L_0}{\left( {{d \over {{d_0}}}} \right)^{ - \beta }}$, where $L_0$ denotes the channel gain at reference distance ${{d_0}}=1\rm m$, $\beta$ is the path loss exponent. We set the path loss exponents for the BS-IRS link, IRS-user link, and BS-user link as $\beta_{br}=2.2$, $\beta_{ru}=2.2$, and $\beta_{bu}=3.6$.   We assume that the BS-user link  follows Rayleigh fading, and BS-IRS link and IRS-user link follow Rician fading with Rician factor $10\rm dB$. Unless otherwise specified,  $N_t=8$, $M=60$, ${\sigma ^2} =  - 80{\rm{dBm}}$, $P_{\rm max}=1\rm W$.
\begin{figure*}
\begin{minipage}[t]{0.33\linewidth}
\centering
\includegraphics[width=2.3in]{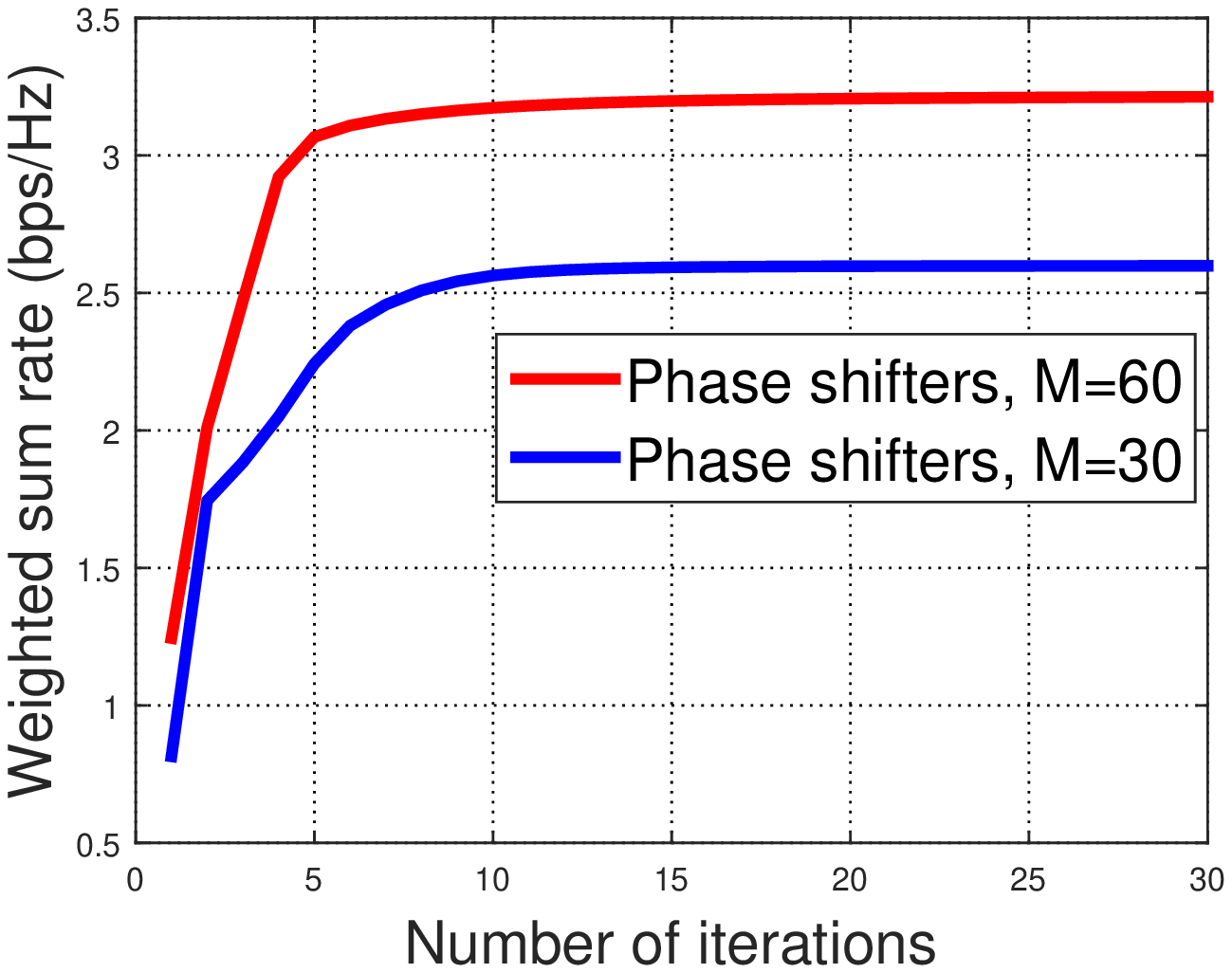}
 \centerline{\scriptsize Fig. 2.  Convergence behavior of the proposed method . }
\end{minipage}%
\begin{minipage}[t]{0.33\linewidth}
\centering
\includegraphics[width=2.3in]{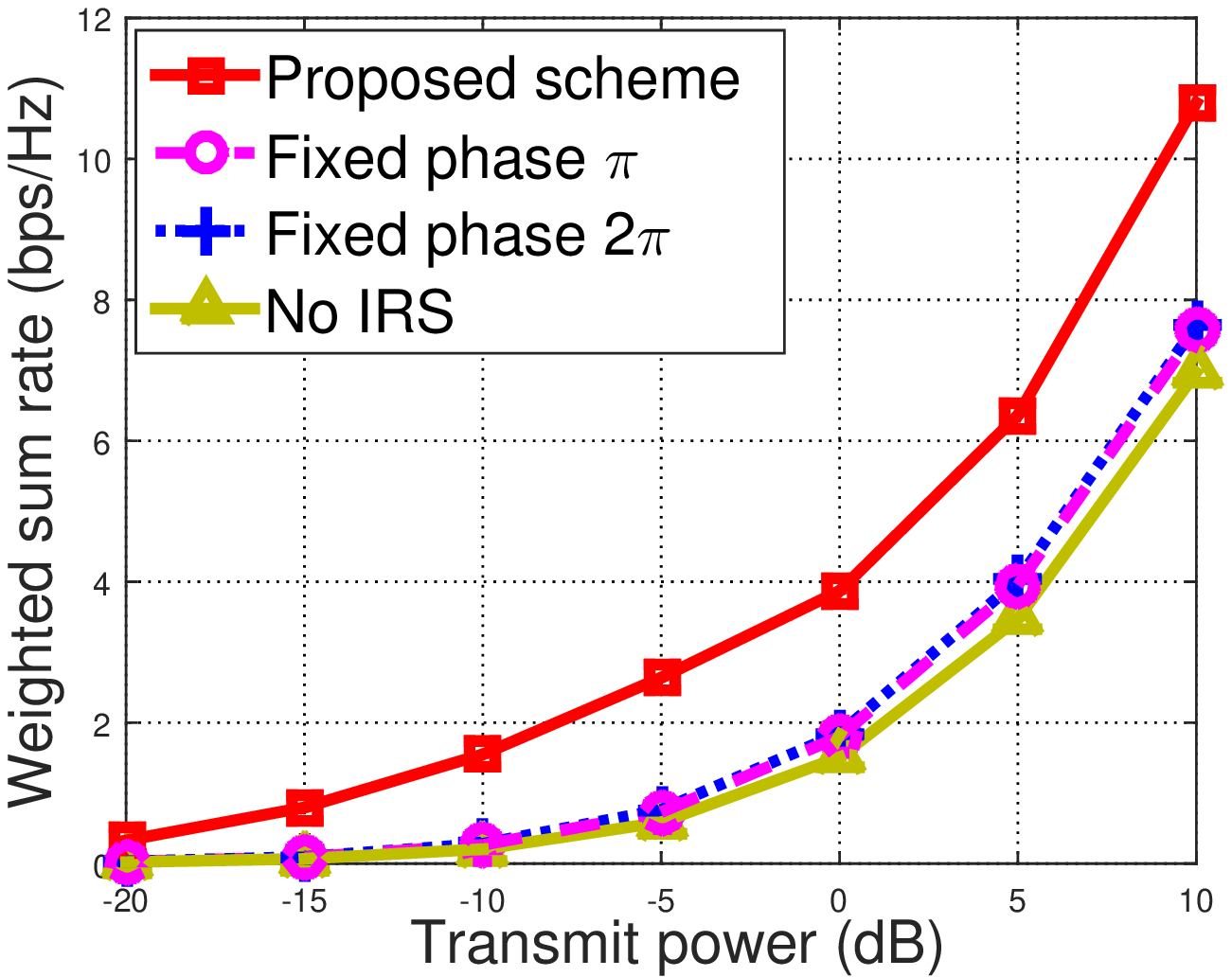}
\centerline{\scriptsize  Fig. 3. WSR versus transmit power.}
\end{minipage}
\begin{minipage}[t]{0.33\linewidth}
\centering
\includegraphics[width=2.3in]{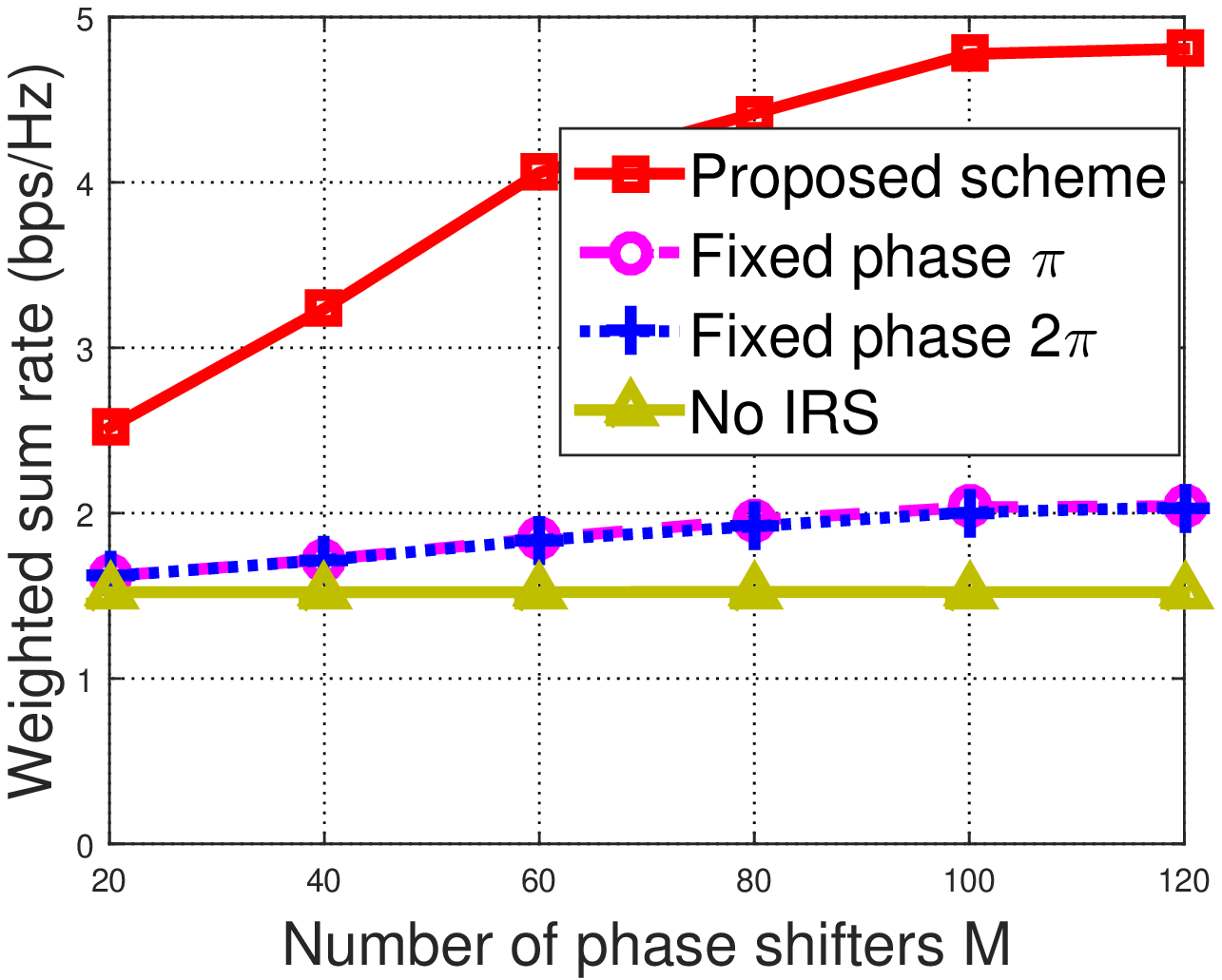}
\centerline{\scriptsize  Fig. 3. WSR versus number of phase shifters $M$.}
\end{minipage}
\end{figure*}

Before evaluating the system performance of our proposed method, the convergence behaviors for different number of phase shifters $M$ are plotted in  Fig.~2. It is observed that the weighted sum rate  monotonically increases with the number of iterations and finally converges within a few iteration. Even when the number of phase shifters reaches $M=60$, the proposed algorithm still has fast convergence behavior.

In Fig.~3, we show  the transmit power versus WSR for different schemes. First, it is observed that the WSR achieved by all the schemes  monotonically increases with the BS's  power limit $P_{\rm max}$. Second, our proposed joint optimization of beamforming and phase shifts scheme outperforms the other benchmarks, especially when the BS's transmit power limit increases, the performance will be more pronounced. Third, compare with no IRS aided communication, the random phase scheme nearly has same performance. This results show that the IRS's  phase shifters must be well tuned so as to improve the system performance.

In Fig.~4, the impact of the number of phase shifters on the WSR has been investigated. First, it can be seen that all the schemes except `No IRS' scheme increases with number of phase shifters $M$. It is expected since more phase shifters will aggregate more signal power to the users, thereby improving  the throughput. Additionally, it also can be seen that the random  phase scheme will not benefit from the number of phase shifters, which again indicates that the important of the optimization of phase shift matrix.

\section{Conclusion}
In this paper, we have studied multiple IRSs aided wireless communications. We formulated the problem as a weighted sum rate  optimization  problem, and a sub-optimal solution of formulated problem was obtained based on the Lagrangian method and  Riemannian manifold  conjugate gradient  method. Numerical results showed that our proposed joint optimization  of  BS transmit beamforming and IRS's phase shifts achieved significantly higher throughput than the  other benchmarks. In addition, the proposed iterative algorithm was quite efficient, which only  requires a  few number of iterations .

\bibliographystyle{IEEEtran}
\bibliography{MultiIRS}
\end{document}